\documentstyle[prd,preprint,tighten,aps,eqsecnum,
amstex,righttag,ctagsplt,epsfig]{revtex}
\begin{document}
\draft
\preprint{
\begin{tabular}{r}
DFTT 48/95
\\
hep-ph/9508277
\end{tabular}
}
\title{ELASTIC $\boldsymbol{\nu}\boldsymbol{N}$ AND
$\boldsymbol{\bar\nu}\boldsymbol{N}$ SCATTERING
AND STRANGE FORM FACTORS OF THE NUCLEONS}
\author{
W.M. Alberico$^{\mathrm{a}}$,
S.M. Bilenky$^{\mathrm{a,b}}$,
C. Giunti$^{\mathrm{a}}$
and
C. Maieron$^{\mathrm{a}}$
}
\address{
\begin{tabular}{c}
$^{\mathrm{a}}$INFN, Sezione di Torino
and Dipartimento di Fisica Teorica, Universit\`a di Torino,
\\
Via P. Giuria 1, 10125 Torino, Italy
\\
$^{\mathrm{b}}$Joint Institute for Nuclear Research, Dubna, Russia
\end{tabular}
}
\date{\today}
\maketitle
\begin{abstract}
Elastic scattering of neutrinos and antineutrinos
on nucleons is considered. It is shown that the
measurement of the neutrino-antineutrino asymmetry
would allow to obtain model independent informations
on the strange axial and vector form factors.
The ratio
of the magnetic form factors of the proton and neutron,
on which the asymmetry depends,
is investigated in detail.
The asymmetry is calculated under different assumptions
on the behavior of the strange vector and axial form factors.
\end{abstract}


\narrowtext

\section{Introduction}
\label{INTRO}

The measurement
performed by the EMC collaboration
\cite{EMC}
of the polarized structure function
of the proton $g_1^p$
led to a big progress
in the investigation of the structure of the nucleon
(see Ref.\cite{ANSELMINO}).
One of the conclusions based on the EMC data
was that the value of the one-nucleon matrix element
of the strange axial current
is comparable with those
of the $u$ and $d$ axial currents.
The latest experiments done at
CERN
\cite{CERN}
and
SLAC
\cite{SLAC}
confirm this conclusion.

The one-nucleon matrix element of the axial
quark current has the following form:
\begin{equation}
\left\langle
p , s
\left|
\bar{q} \gamma^{\alpha} \gamma_{5} q
\right|
p , s
\right\rangle
=
2 M s^{\alpha} g_{A}^{q}
\;.
\label{E001}
\end{equation}
Here $p$ is the nucleon momentum,
$M$ is the nucleon mass,
$s^{\alpha}$ is the spin vector
and
$g_{A}^{q}$ is a constant.

The constants
$g_{A}^{u}$,
$g_{A}^{d}$
and
$g_{A}^{s}$
can be determined from the following three constraints:

\begin{enumerate}

\item
The QCD sum rule
\cite{QCDSUMRULE,ANSELMINO}
\begin{equation}
\Gamma_{1}^{p}
=
\int_{0}^{1}
g_{1}^{p}
\,
\mathrm{d} x
=
{\displaystyle1\over\displaystyle2}
\left(
{\displaystyle4\over\displaystyle9}
\,
g_{A}^{u}
+
{\displaystyle1\over\displaystyle9}
\,
g_{A}^{d}
+
{\displaystyle1\over\displaystyle9}
\,
g_{A}^{s}
\right)
\left(
1
-
{\displaystyle
\alpha_{s}
\over\displaystyle
\pi
}
+
\ldots
\right)
\;,
\label{E002}
\end{equation}
Here
$ \alpha_{s} $
is the QCD coupling constant.

\item
The relation
\begin{equation}
g_{A}
=
g_{A}^{u}
-
g_{A}^{d}
\;,
\label{E003}
\end{equation}
where the constant
$ g_{A} = 1.2573 \pm 0.0028 $
is the axial constant,
which
is determined from neutron decay.

\item
The relation
\begin{equation}
3 F - D
=
g_{A}^{u}
+
g_{A}^{d}
-
2
g_{A}^{s}
\;.
\label{E004}
\end{equation}
The values of the constants $F$ and $D$ can be determined from
semileptonic decays of hyperons.
{}From the latest data
it follows
\cite{CLOSE}:
$ F = 0.459 \pm 0.008 $
and
$ D = 0.798 \pm 0.008 $.

\end{enumerate}

The relation (\ref{E003})
is based on isotopic SU(2) invariance.
The relation (\ref{E004})
is derived from exact SU(3) invariance.

{}From the data of the latest experiments
\cite{CERN,SLAC}
on the measurement of $g_1^p$ it has been found that
\begin{equation}
\arraycolsep=0cm
\begin{array}{lcl} \displaystyle
\Gamma_{1}^{p}
=
0.133 \pm 0.04 \pm 0.012
\null & \null \hspace{1cm} \null & \null \displaystyle
\mbox{(E143)}
\;,
\\ \displaystyle
\Gamma_{1}^{p}
=
0.136 \pm 0.011 \pm 0.011
\null & \null \hspace{1cm} \null & \null \displaystyle
\mbox{(SMC)}
\;.
\end{array}
\label{E005}
\end{equation}

The global fit of all data gives
\cite{Ellis}
\begin{equation}
\begin{array}{l} \displaystyle
g_{A}^{u}
=
0.83 \pm 0.03
\,
\\ \displaystyle
g_{A}^{d}
=
- 0.43 \pm 0.03
\,
\\ \displaystyle
g_{A}^{s}
=
- 0.10 \pm 0.03
\,
\end{array}
\label{E006}
\end{equation}

Thus,
the value of the constant $g_{A}^{s}$,
that characterizes
the matrix element of the strange axial current,
is comparable with those of the constants
$ g_{A}^{u} $
and
$ g_{A}^{d} $.
This fact has brought a lot of discussions in the literature
(see Ref.\cite{ANSELMINO}).

In the parton model the constants $ g_{A}^{q} $ are easily
calculated.
The result is well known:
\begin{equation}
g_{A}^{q}
=
\Delta q
\equiv
\int_{0}^{1}
\left(
\sum_{r} r q^{r}(x)
+
\sum_{r} r \bar{q}^{r}(x)
\right)
\mathrm{d} x
\;,
\label{E007}
\end{equation}
where
$ q^{r}(x) $
($ \bar{q}^{r}(x) $)
is the density of quarks $q$
(antiquarks $\bar{q}$)
with momentum fraction $x$
and helicity $r$.

Thus,
in the parton model,
the constant
$ g_{A}^{q} $
is the contribution of $q$ (and $\bar{q}$)
to the helicity of proton.
In the parton model
the total contribution of the $u$, $d$ and $s$
quarks (and antiquarks)
to the helicity of the proton
is given by
\begin{equation}
\sum_{q=u,d,s} \Delta q
=
\sum_{q=u,d,s} g_{A}^{q}
\;.
\label{E008}
\end{equation}

{}From the analysis of the EMC data it was obtained that
the averaged value of the constant
$ \displaystyle
\sum_{q=u,d,s} g_{A}^{q}
$
is close to zero.
That was the origin
of the ``problem of the spin of the proton''.
One way of solving this problem was the calculation
of the contribution of gluons to the helicity of the proton.
It was shown
\cite{ANOMALY}
that,
due to the triangle anomaly,
this contribution can be large.
According to the latest data
$ \displaystyle
\sum_{q=u,d,s} g_{A}^{q}
=
0.31 \pm 0.07
$.

The values
of the constants $g_{A}^{q}$
given in Eq.(\ref{E006})
were obtained
under several assumptions.
The calculation of the quantity
$ \Gamma_{1}^{p} $
that enter in the
QCD sum rule (\ref{E002})
requires the extrapolation of the experimental data
in the region of small $x$.
The value of the integral
$ \Gamma_{1}^{p} $
depends significantly on
the assumptions on the behaviour
of the function $g_{1}^{p}(x)$ at small $x$.
In some models
(see, for example, Refs.\cite{Ellis,Landshoff})
this
function can have a singularity at $x=0$.
Another important assumption is the exact SU(3) symmetry.
It was shown in Ref.\cite{SU3}
that a violation
of the SU(3) symmetry can change strongly the numerical value
of the constant $g_{A}^{s}$.

It is then clear that it is very important to use other methods
for the determination of the matrix elements of the strange current.
The investigation of
neutral-current (NC) neutrino
reactions is one of these ways
\cite{NEUTRINOP}.
We will consider here the elastic scattering of muon neutrinos
and antineutrinos on nucleons:
\begin{eqnarray}
&&
\nu_\mu + N \to \nu_\mu + N
\;,
\label{E009}
\\
&&
\bar\nu_\mu + N \to \bar\nu_\mu + N
\;.
\label{E010}
\end{eqnarray}

The cross sections of these processes depend
on the electromagnetic form factors,
on the axial form factor and on the strange axial and vector form factors
of the nucleon.
As it was shown in Ref.\cite{Garvey},
the value of the constant $g_{A}^{s}$
that can be extracted from the cross sections of the processes
(\ref{E009}) and (\ref{E010})
strongly depends on the behaviour of the axial form factor,
which is rather poorly known.
In order to minimize this dependence
we will consider here the asymmetry
\begin{equation}
\cal{A}_{N}(Q^2)
=
{\displaystyle
\left(
{\displaystyle
\mathrm{d} \sigma
\over\displaystyle
\mathrm{d} Q^2
}
\right)_{\nu N}^{\mathrm{NC}}
-
\left(
{\displaystyle
\mathrm{d} \sigma
\over\displaystyle
\mathrm{d} Q^2
}
\right)_{\bar\nu N}^{\mathrm{NC}}
\over\displaystyle
\left(
{\displaystyle
\mathrm{d} \sigma
\over\displaystyle
\mathrm{d} Q^2
}
\right)_{\nu n}^{\mathrm{CC}}
-
\left(
{\displaystyle
\mathrm{d} \sigma
\over\displaystyle
\mathrm{d} Q^2
}
\right)_{\bar\nu p}^{\mathrm{CC}}
}
\;,
\label{E011}
\end{equation}
where
$ \displaystyle
\left(
{\displaystyle
\mathrm{d} \sigma
\over\displaystyle
\mathrm{d} Q^2
}
\right)_{\nu N}^{\mathrm{NC}}
$
and
$ \displaystyle
\left(
{\displaystyle
\mathrm{d} \sigma
\over\displaystyle
\mathrm{d} Q^2
}
\right)_{\bar\nu N}^{\mathrm{NC}}
$
are the cross sections of the
processes (\ref{E009}) and (\ref{E010}),
respectively,
while
$ \displaystyle
\left(
{\displaystyle
\mathrm{d} \sigma
\over\displaystyle
\mathrm{d} Q^2
}
\right)_{\nu n}^{\mathrm{CC}}
$
and
$ \displaystyle
\left(
{\displaystyle
\mathrm{d} \sigma
\over\displaystyle
\mathrm{d} Q^2
}
\right)_{\bar\nu p}^{\mathrm{CC}}
$
are,
respectively,
the cross sections of the quasi-elastic
charged-current (CC) processes
\begin{eqnarray}
&&
\nu_\mu + n \to \mu^{-} + p
\;,
\label{E012}
\\
&&
\bar\nu_\mu + p \to \mu^{+} + n
\;.
\label{E013}
\end{eqnarray}
Here
$ Q^2 \equiv - q^2 $,
where
$ q = p' - p $,
$p$ and $p'$
are the momenta of the initial and final nucleons,
respectively.

We will show that measurements of the asymmetry
$ \cal{A}_{N}(Q^2) $
will allow to obtain model independent informations
on the strange axial and vector form factors of the nucleon.

Let us conclude this introduction
with the following remark.
The last experiment on the measurement of the cross sections
of elastic NC scattering of neutrinos and antineutrinos on protons
was done about 10 years ago in BNL \cite{Ahrens}.
We think that now it is a proper time to continue
these investigations.
A wide program of long-baseline
neutrino experiments are under preparation and under discussion at the
moment \cite{LBNO}.
The main aim of these experiments is
to search for neutrino oscillations in a distant detector.
A detector will be placed near the neutrino source
for the calibration of the beam.
This detector
will register a large number of
neutrino events.
Taking into
account the importance of the knowledge of the strange
form factors of the nucleon,
we think that
the possibility to perform a
measurement of the cross section of elastic scattering
of neutrinos and antineutrinos on
nucleons in the ``near'' detector
should be considered seriously.

\section{The Asymmetry}
\label{ASYMMETRY}

Let us consider the elastic scattering of
$\nu_\mu$ and $\bar\nu_\mu$
on nucleons.
The standard effective Hamiltonian for these processes is given by
\begin{equation}
\cal{H}_{I}^{\mathrm{NC}}
=
{\displaystyle
G
\over\displaystyle
\sqrt{2}
}
\,
\bar\nu_\mu
\gamma^{\alpha}
\left( 1 + \gamma_{5} \right)
\nu_\mu
j^Z_{\alpha}
\;,
\label{E014}
\end{equation}
where the neutral current
$j^Z_{\alpha}$
can be written in the form
\begin{equation}
j^Z_{\alpha}
=
v^{3}_{\alpha}
+
a^{3}_{\alpha}
-
2 \sin^2\theta_{W} j^{\mathrm{em}}_{\alpha}
-
{\displaystyle1\over\displaystyle2}
\,
v^{s}_{\alpha}
-
{\displaystyle1\over\displaystyle2}
\,
a^{s}_{\alpha}
\label{E015}
\end{equation}
Here
$j^{\mathrm{em}}_{\alpha}$
is the electromagnetic current,
$v^{3}_{\alpha}$
and
$a^{3}_{\alpha}$
are the third
components of the isovector vector and axial currents
\begin{eqnarray}
&&
v^{i}_{\alpha}
=
\bar{N} \gamma_{\alpha}
\,
{\displaystyle\tau^{i}\over\displaystyle2}
\,
N
\;,
\label{E016}
\\
&&
a^{i}_{\alpha}
=
\bar{N} \gamma_{\alpha} \gamma_{5}
\,
{\displaystyle\tau^{i}\over\displaystyle2}
\,
N
\;,
\label{E017}
\end{eqnarray}
($ \displaystyle
N
=
\left(
\begin{array}{l} \displaystyle
u
\\ \displaystyle
d
\end{array}
\right)
$
is the SU(2) doublet)
and
\begin{eqnarray}
&&
v^{s}_{\alpha}
=
\bar{s} \gamma_{\alpha} s
\;,
\label{E018}
\\
&&
a^{s}_{\alpha}
=
\bar{s} \gamma_{\alpha} \gamma_{5} s
\label{E019}
\end{eqnarray}
are the strange vector and axial currents.
In Eq.(\ref{E015}) we kept only the contribution to
the neutral current of the light $u$, $d$ and $s$ quarks.

The one-nucleon matrix element of the neutral current has the
following general form
\begin{equation}
\left\langle
p'
\left|
J^Z_{\alpha}
\right|
p
\right\rangle
=
\bar{u}(p')
\left[
\gamma_{\alpha}
F_V^Z(Q^2)
+
{\displaystyle
i
\over\displaystyle
2 M
}
\,
\sigma_{\alpha\beta}
q^{\beta}
F_M^Z(Q^2)
+
\gamma_{\alpha}
\gamma_5
F_A^Z(Q^2)
\right]
u(p)
\;.
\label{E020}
\end{equation}

{}From Eq.(\ref{E015}),
for the vector NC
form factors we have
\begin{eqnarray}
&&
(F_V^Z)_{p(n)}
=
\pm
F_1^3
-
2 \sin^2\theta_W F_1^{p(n)}
-
{\displaystyle1\over\displaystyle2}
F_V^s
\;,
\label{E021}
\\
&&
(F_M^Z)_{p(n)}
=
\pm
F_2^3
-
2 \sin^2\theta_W F_2^{p(n)}
-
{\displaystyle1\over\displaystyle2}
F_M^s
\;.
\label{E022}
\end{eqnarray}
Here
$ F_1^{p(n)} $
and
$ F_2^{p(n)} $
are the Dirac and Pauli electromagnetic
form factors of the proton (neutron),
$F_V^s$
and
$F_M^s$
are the strange
vector form factors
and
$F_{1,2}^3$
are the isovector form factors of the nucleon.
{}From isotopic invariance it follows that
\begin{equation}
F_{1,2}^3
=
{\displaystyle1\over\displaystyle2}
\left(
F_{1,2}^p
-
F_{1,2}^n
\right)
\label{E023}
\end{equation}

Furthermore,
using the isotopic SU(2) invariance of strong interactions
we have
\begin{equation}
\sideset{_{p}}{_{p}}
{
\left\langle
p'
\left|
A_{\alpha}^{3}
\right|
p
\right\rangle
}
=
-
\sideset{_{n}}{_{n}}
{
\left\langle
p'
\left|
A_{\alpha}^{3}
\right|
p
\right\rangle
}
=
{\displaystyle1\over\displaystyle2}
\,
\sideset{_{p}}{_{n}}
{
\left\langle
p'
\left|
A_{\alpha}^{1+i2}
\right|
p
\right\rangle
}
\;,
\label{E024}
\end{equation}
$
A_{\alpha}^{1+i2}
=
A_{\alpha}^{1}
+
i A_{\alpha}^{2}
$
being the charged axial current.
{}From Eq.(\ref{E015}) we obtain
\begin{equation}
(F_A^Z)_{p(n)}
=
\pm
{\displaystyle1\over\displaystyle2}
F_A
-
{\displaystyle1\over\displaystyle2}
F_A^s
\;.
\label{E025}
\end{equation}
where
$ F_A $
is the CC axial form factor
and
$ F_A^s $
is the axial strange form factor.

Using Eqs.(\ref{E021}), (\ref{E022}) and (\ref{E025}),
for the difference of
cross sections of the NC processes
(\ref{E009}) and (\ref{E010})
we obtain the following expression:
\begin{equation}
\arraycolsep=0cm
\begin{array}{rcl} \displaystyle
\left(
{\displaystyle
\mathrm{d} \sigma
\over\displaystyle
\mathrm{d} Q^2
}
\right)^{\mathrm{NC}}_{\nu p(n)}
-
\left(
{\displaystyle
\mathrm{d} \sigma
\over\displaystyle
\mathrm{d} Q^2
}
\right)^{\mathrm{NC}}_{\bar\nu p(n)}
\null & \null = \null & \null \displaystyle
{\displaystyle
G^2
\over\displaystyle
\pi
}
\,
{\displaystyle
Q^2
\over\displaystyle
p \cdot k
}
\left(
1
-
{\displaystyle
Q^2
\over\displaystyle
4 \, p \cdot k
}
\right)
\\ \displaystyle
\null & \null \times \null & \null \displaystyle
\left(
\pm
G_M^3
-
2 \sin^2\theta_W G_M^{p(n)}
-
{\displaystyle1\over\displaystyle2}
G_M^s
\right)
\left(
\pm
{\displaystyle1\over\displaystyle2}
F_A
-
{\displaystyle1\over\displaystyle2}
F^s_A
\right)
\;,
\end{array}
\label{E026}
\end{equation}
where $k$ is the momentum of the initial neutrino,
$
G_M^{p(n)}
=
F_1^{p(n)}
+
F_2^{p(n)}
$
is magnetic form factor of the proton (neutron),
$
G_M^3
=
{1\over2}
\left(
G_M^p
-
G_M^n
\right)
$
is the isovector magnetic form factor
and
$
G_M^s
=
F_V^s
+
F_M^s
$
is the strange magnetic form factor.

Let us consider now the quasi-elastic CC processes
(\ref{E012}) and (\ref{E013}).
The standard effective Hamiltonian for these
processes is given by
\begin{equation}
\cal{H}_{I}^{\mathrm{CC}}
=
{\displaystyle
G
\over\displaystyle
\sqrt{2}
}
\,
\bar\mu
\gamma^{\alpha}
\left( 1 + \gamma_{5} \right)
\nu_\mu
j_{\alpha}
+
\mbox{h.c.}
\;.
\label{E027}
\end{equation}
Here
\begin{equation}
j_{\alpha}
=
\left[
\bar{u}
\gamma^{\alpha}
\left( 1 + \gamma_{5} \right)
d
\right]
V_{ud}
=
\left[
v_{\alpha}^{1+i2}
+
a_{\alpha}^{1+i2}
\right]
V_{ud}
\label{E028}
\end{equation}
is the hadronic charged current and
$ V_{ud} $
is the element of C-K-M mixing matrix.
The one-nucleon matrix element
of the CC has the form
\begin{equation}
\arraycolsep=0cm
\begin{array}{rcl} \displaystyle
\sideset{_{p}}{_{n}}
{
\left\langle
p'
\left|
J_{\alpha}^{1+i2}
\right|
p
\right\rangle
}
\null & \null = \null & \null \displaystyle
\sideset{_{n}}{_{p}}
{
\left\langle
p'
\left|
J_{\alpha}^{1-i2}
\right|
p
\right\rangle
}
\\ \displaystyle
\null & \null = \null & \null \displaystyle
\bar{u}(p')
\left[
\gamma_{\alpha}
F_V(Q^2)
+
{\displaystyle
i
\over\displaystyle
2 M
}
\,
\sigma_{\alpha\beta}
q^{\beta}
F_M(Q^2)
+
\gamma_{\alpha}
\gamma_5
F_A(Q^2)
\right]
u(p)
\;.
\end{array}
\label{E029}
\end{equation}
{}From isotopic SU(2) invariance it follows that
\begin{equation}
\begin{array}{l} \displaystyle
F_V
=
F_1^p
-
F_1^n
=
2
\,
F_1^3
\\ \displaystyle
F_M
=
F_2^p
-
F_2^n
=
2
\,
F_2^3
\end{array}
\label{E030}
\end{equation}

Using Eq.(\ref{E030}),
for the difference of cross sections of the CC processes
(\ref{E012}) and (\ref{E013})
we obtain
\begin{equation}
\left(
{\displaystyle
\mathrm{d} \sigma
\over\displaystyle
\mathrm{d} Q^2
}
\right)_{\nu n}^{\mathrm{CC}}
-
\left(
{\displaystyle
\mathrm{d} \sigma
\over\displaystyle
\mathrm{d} Q^2
}
\right)_{\bar\nu p}^{\mathrm{CC}}
=
{\displaystyle
G^2
\over\displaystyle
\pi
}
\,
{\displaystyle
Q^2
\over\displaystyle
p \cdot k
}
\left(
1
-
{\displaystyle
Q^2
\over\displaystyle
4 \, p \cdot k
}
\right)
2
\,
G_M^3
\,
F_A
\left| V_{ud} \right|^2
\label{E031}
\end{equation}
Finally,
from Eqs.(\ref{E026}) and (\ref{E031}),
for the asymmetries $ \cal{A}_{p(n)} $
determined by the expression (\ref{E011})
we find the following expression:
\begin{equation}
\cal{A}_{p(n)}
=
{\displaystyle
1
\over\displaystyle
4 \left| V_{ud} \right|^2
}
\left(
\pm
1
-
{\displaystyle
F^s_A
\over\displaystyle
F_A
}
\right)
\left(
\pm
1
-
2 \sin^2\theta_W
{\displaystyle
G_M^{p(n)}
\over\displaystyle
G_M^3
}
-
{\displaystyle
G_M^s
\over\displaystyle
2 \, G_M^3
}
\right)
\;.
\label{E032}
\end{equation}
Taking into account only the terms which
depend linearly on the strange form factors,
we can rewrite Eq.(\ref{E032}) in the form
\begin{equation}
R_{p(n)}(Q^2)
\,
\cal{A}_{p(n)}(Q^2)
=
1
\mp
{\displaystyle
F^s_A(Q^2)
\over\displaystyle
F_A(Q^2)
}
\mp
{\displaystyle
1
\over\displaystyle
8 \left| V_{ud} \right|^2
}
\,
R_{p(n)}(Q^2)
\,
{\displaystyle
G_M^s(Q^2)
\over\displaystyle
G_M^3(Q^2)
}
\;,
\label{E033}
\end{equation}
where the quantity
\begin{equation}
R_{p(n)}(Q^2)
=
{\displaystyle
4 \left| V_{ud} \right|^2
\over\displaystyle
1 \mp 2 \sin^2\theta_W
{\displaystyle
G_M^{p(n)}(Q^2)
\over\displaystyle
G_M^3(Q^2)
}
}
\label{E034}
\end{equation}
is determined by the electromagnetic form factors
of the nucleon.
Thus,
the measurement of the asymmetry
$ \cal{A}_{p(n)}(Q^2) $
would allow to obtain informations
about the strange axial and vector
form factors
directly from the experimental data.
If it will turn out that
the left-hand side of Eq.(\ref{E033}),
in which only measurable quantities enter,
is different from 1,
it will be a model independent proof
that the strange nucleon form factors are different
from zero.

In neutrino experiments usually
the interactions of neutrinos with nuclei
are detected.
{}From Eq.(\ref{E026}),
for the asymmetry
averaged over the proton and neutron
we obtain the following relation:
\begin{equation}
r
\,
\cal{A}(Q^2)
=
1
+
{\displaystyle
r \, \sin^2\theta_W
\over\displaystyle
2 \left| V_{ud} \right|^2
}
\,
{\displaystyle
G_M^0(Q^2)
\over\displaystyle
G_M^3(Q^2)
}
\,
{\displaystyle
F^s_A(Q^2)
\over\displaystyle
F_A(Q^2)
}
\;,
\label{E035}
\end{equation}
where
\begin{equation}
r
=
{\displaystyle
4 \left| V_{ud} \right|^2
\over\displaystyle
1 - 2 \sin^2\theta_W
}
\label{E036}
\end{equation}
and
$ \displaystyle
G_M^0
=
{\displaystyle1\over\displaystyle2}
\left(
G_M^p
+
G_M^n
\right)
$
is the isoscalar magnetic form factor of the nucleon.
It is obvious that the interference
of the isoscalar strange vector form factor
$ G_M^s $
and the isovector axial form factor
$ F_A $
vanishes after averaging over $p$ and $n$.
Thus,
the averaged asymmetry
$ \cal{A}(Q^2) $
is determined only by the axial strange form factor
$ F_A^s $.

The electromagnetic form factors of the nucleon
enter into the expression for the asymmetry
$ \cal{A}_{p(n)}(Q^2) $
in the form of the ratio
$ G_M^{p(n)} / G_M^3 $.
As it is well known,
the electromagnetic form factors
satisfy the approximate scaling relations
\begin{equation}
\begin{array}{l} \displaystyle
G_M^p(Q^2)
=
\mu_p
\,
G_E^p(Q^2)
\;,
\\ \displaystyle
G_M^n(Q^2)
=
\mu_n
\,
G_E^p(Q^2)
\;,
\end{array}
\label{E037}
\end{equation}
where
$ \mu_{p(n)} $
is the total magnetic moment
of the proton (neutron) in nuclear Bohr magnetons.

Using the values
\cite{RPP}
\begin{equation}
\begin{array}{lcl} \displaystyle
\mu_p
=
2.793
\;,
\null & \null \qquad \null & \null \displaystyle
\mu_n
=
-
1.913
\;,
\\ \displaystyle
\null & \null \qquad \null & \null \displaystyle
\\ \displaystyle
\sin^2\theta_W
=
0.232
\;,
\null & \null \qquad \null & \null \displaystyle
\left| V_{ud} \right|
=
0.975
\;,
\end{array}
\label{E038}
\end{equation}
for the asymmetries
$ \cal{A}_{p} $,
$ \cal{A}_{n} $
and the averaged asymmetry $ \cal{A} $
we obtain the following expressions
in the scaling approximation:
\begin{eqnarray}
&&
R_{p}
\,
\cal{A}_{p}(Q^2)
=
1
-
{\displaystyle
F^s_A(Q^2)
\over\displaystyle
F_A(Q^2)
}
-
1.11
\,
{\displaystyle
G_M^s(Q^2)
\over\displaystyle
G_M^3(Q^2)
}
\;,
\label{E039}
\\
&&
R_{n}
\,
\cal{A}_{n}(Q^2)
=
1
+
{\displaystyle
F^s_A(Q^2)
\over\displaystyle
F_A(Q^2)
}
+
0.80
\,
{\displaystyle
G_M^s(Q^2)
\over\displaystyle
G_M^3(Q^2)
}
\;,
\label{E040}
\\
&&
r
\,
\cal{A}(Q^2)
=
1
+
0.16
\,
{\displaystyle
F^s_A(Q^2)
\over\displaystyle
F_A(Q^2)
}
\;,
\label{E041}
\end{eqnarray}
where
$ R_p = 8.46 $,
$ R_n = 6.11 $
and
$ r = 7.09 $.
Thus,
the asymmetries $ \cal{A}_{p(n)} $ are rather sensitive to
the axial and vector strange form factors.
It is seen from
Eqs.(\ref{E039}) and (\ref{E040})
that the asymmetries $ \cal{A}_{p(n)}(Q^2) $
have approximately the same sensitivity
to the axial and vector strange form factors.
The contribution of the axial strange form factor
to the averaged asymmetry $ \cal{A}(Q^2) $
is suppressed due to the
smallness of the isoscalar part
of the magnetic moment of the nucleon
with respect to its isovector part.

We calculated also the integral asymmetry
\begin{equation}
\Bbb{A}_{N}
=
{\displaystyle
\sigma_{\nu N}^{\mathrm{NC}}
-
\sigma_{\bar\nu N}^{\mathrm{NC}}
\over\displaystyle
\sigma_{\nu n}^{\mathrm{CC}}
-
\sigma_{\bar\nu p}^{\mathrm{CC}}
}
\;,
\label{E0421}
\end{equation}
where
$ \sigma_{\nu N}^{NC}$,
$ \sigma_{\bar{\nu} N}^{NC}$,
$\sigma_{\nu n}^{CC}$ and
$\sigma_{\bar{\nu} p}^{CC}$
are the total cross sections
of the processes
(\ref{E009}), (\ref{E010}), (\ref{E012}) and (\ref{E013}),
respectively.
{}From Eqs.(\ref{E026}) and (\ref{E031})
we have
\begin{equation}
\arraycolsep=0cm
\begin{array}{rcl} \displaystyle
\Bbb{A}_{p(n)}
\null & \null = \null & \null \displaystyle
{\displaystyle
1
\over\displaystyle
4 \left| V_{ud} \right|^2
}
\left[
1
\mp
2 \sin^2\theta_W
{\displaystyle
\int \mathrm{d} Q^2
\,
X(Q^2)
\,
F_A(Q^2)
\,
G_M^{p(n)}(Q^2)
\over\displaystyle
\int \mathrm{d} Q^2
\,
X(Q^2)
\,
F_A(Q^2)
\,
G_M^3(Q^2)
}
\right.
\\ \displaystyle
\null & \null \null & \null \displaystyle
\phantom{
{\displaystyle
1
\over\displaystyle
4 \left| V_{ud} \right|^2
}
\left[
1
\right.
}
\left.
\mp
{\displaystyle
\int \mathrm{d} Q^2
\,
X(Q^2)
\,
F_A^s(Q^2)
\,
G_M^3(Q^2)
\left(
1
\mp
2 \sin^2\theta_W
\,
{\displaystyle
G_M^{p(n)}(Q^2)
\over\displaystyle
G_M^3(Q^2)
}
\right)
\over\displaystyle
\int \mathrm{d} Q^2
\,
X(Q^2)
\,
F_A(Q^2)
\,
G_M^3(Q^2)
}
\right.
\\ \displaystyle
\null & \null \null & \null \displaystyle
\phantom{
{\displaystyle
1
\over\displaystyle
4 \left| V_{ud} \right|^2
}
\left[
1
\right.
}
\left.
\mp
{1\over2}
\,
{\displaystyle
\int \mathrm{d} Q^2
\,
X(Q^2)
\,
F_A(Q^2)
\,
G_M^s(Q^2)
\over\displaystyle
\int \mathrm{d} Q^2
\,
X(Q^2)
\,
F_A(Q^2)
\,
G_M^3(Q^2)
}
\right]
\;,
\end{array}
\label{E042}
\end{equation}
where
\begin{equation}
X(Q^2)
=
{\displaystyle
Q^2
\over\displaystyle
p \cdot k
}
\left(
1
-
{\displaystyle
Q^2
\over\displaystyle
4 \, p \cdot k
}
\right)
\label{E043}
\end{equation}

Let us stress that the integral asymmetry depends
not only on the strange form factors,
but also on the axial and electromagnetic form factors.
In the scaling approximation,
from Eq.(\ref{E042}) for the integral
asymmetries we obtain
\begin{equation}
\arraycolsep=0cm
\begin{array}{rcl} \displaystyle
R_{p(n)}
\,
\Bbb{A}_{p(n)}
\null & \null = \null & \null \displaystyle
1
\mp
{\displaystyle
\int \mathrm{d} Q^2
\,
X(Q^2)
\,
F_A^s(Q^2)
\,
G_M^3(Q^2)
\over\displaystyle
\int \mathrm{d} Q^2
\,
X(Q^2)
\,
F_A(Q^2)
\,
G_M^3(Q^2)
}
\\ \displaystyle
\null & \null \null & \null \displaystyle
\phantom{ 1 }
\mp
{\displaystyle
R_{p(n)}
\over\displaystyle
8 \left| V_{ud} \right|^2
}
\,
{\displaystyle
\int \mathrm{d} Q^2
\,
X(Q^2)
\,
F_A(Q^2)
\,
G_M^s(Q^2)
\over\displaystyle
\int \mathrm{d} Q^2
\,
X(Q^2)
\,
F_A(Q^2)
\,
G_M^3(Q^2)
}
\,
\end{array}
\label{E044}
\end{equation}
where
$ R_p / 8 \left| V_{ud} \right|^2 = 1.11 $
and
$ R_n / 8 \left| V_{ud} \right|^2 = 0.80 $.
Thus,
also
the integral asymmetries are very sensitive
to the axial and vector strange form factors.

The integral asymmetry averaged over $p$ and $n$ obviously does not
depend on the vector strange form factor and is given by the expression
\begin{equation}
\Bbb{A}
=
{\displaystyle
1
\over\displaystyle
4 \left| V_{ud} \right|^2
}
\left[
1
-
2 \sin^2\theta_W
+
2 \sin^2\theta_W
\,
{\displaystyle
\int \mathrm{d} Q^2
\,
X(Q^2)
\,
F_A^s(Q^2)
\,
G_M^0(Q^2)
\over\displaystyle
\int \mathrm{d} Q^2
\,
X(Q^2)
\,
F_A(Q^2)
\,
G_M^3(Q^2)
}
\right]
\;.
\label{E045}
\end{equation}
Thus,
if it will turn out that the averaged integral asymmetry
differs from the value
$ \displaystyle
\left( 1 - 2 \sin^2\theta_W \right)
/
4 \left| V_{ud} \right|^2
$,
it will be a model independent proof that
the axial strange form factor is different from zero.

If form factor scaling takes place,
for the averaged integral asymmetry we have
\begin{equation}
\arraycolsep=0cm
\begin{array}{rcl} \displaystyle
r
\,
\Bbb{A}
\null & \null = \null & \null \displaystyle
1
+
0.16
\,
{\displaystyle
\int \mathrm{d} Q^2
\,
X(Q^2)
\,
F_A^s(Q^2)
\,
G_E^p(Q^2)
\over\displaystyle
\int \mathrm{d} Q^2
\,
X(Q^2)
\,
F_A(Q^2)
\,
G_E^p(Q^2)
}
\;.
\end{array}
\label{E046}
\end{equation}

\section{Discussion}
\label{DISCUSSION}

{}From the data of the experiments on deep inelastic
scattering of polarized leptons on polarized
nucleons and from the data of other experiments
it was found that the axial strange constant $g_A^s$
is relatively large.
These data led to an enormous
amount of theoretical investigations and a big progress in
the understanding of the structure of the nucleon
\cite{ANSELMINO}.
Taking into account the importance of
the ``problem of the spin of the proton''
and the theoretical uncertainties connected with
the analysis of the data
(small $x$ extrapolation, violation of SU(3), etc.),
it is very important to use other methods
for the determination of
the matrix elements
of the strange currents.

The investigation of neutrino processes at relatively small $Q^2$
can be an important source of information about
the matrix elements
of the strange axial and vector currents
\cite{NEUTRINOP}.

We have considered the processes of elastic scattering of neutrinos
and antineutrinos on nucleons.
{}From the isotopic SU(2)
invariance of strong interactions
it follows that the cross sections of these
processes are determined by the electromagnetic
form factors of the proton and the neutron,
by the axial CC form factor of the nucleon
and by the strange vector and axial form factors.
In order to minimize
the uncertainties connected with
the knowledge of the electromagnetic
form factors of the neutron and,
especially,
of the axial form factor of the nucleon,
we considered here the neutrino-antineutrino asymmetry
$ \cal{A}_{N} $
determined by the relation (\ref{E011}).
A measurement of the
asymmetry $ \cal{A}_{N} $ would allow to obtain
directly from the experimental data
model independent informations on the
strange vector and axial form factors of the nucleon.

In Section \ref{ASYMMETRY}
we have shown that the asymmetry
$ \cal{A}_{N} $
depends
on the magnetic form factors of the proton and neutron,
which are known with better accuracy than the electric ones.
More specifically,
the asymmetry
$ \cal{A}_{N} $
depends on the ratio of the magnetic form factors of
the neutron and proton.
Rather detailed informations
on the values of the proton form factors
in a wide range of $Q^2$
(up to
$ 30 \, \mbox{GeV}^2 $)
were obtained
from the data on elastic scattering
of electrons on proton
\cite{Data_p,Bartel}.
It was clearly shown that there is
a deviation of the form factors from the dipole formula
(up to 30\% at high $Q^2$).

The values of the neutron form factors are much less known
than those of the proton.
A large part of the data
on the neutron form factors were obtained
from the analysis of quasi-elastic scattering
of electrons on nuclei (in particular deuterium).
This analysis
requires the consideration
of many theoretical effects
(final state interaction,
contributions of meson exchange currents, etc.).
The values of these form factors
in the range
$ 1.75 \le Q^2 \le 4 \, \mbox{GeV}^2 $
were determined
in a recent SLAC experiment
\cite{Lung}.
Some informations on the value of
$ G_M^n $
are also available in the region
$ 4 \le Q^2 \le 10 \, \mbox{GeV}^2 $
\cite{Rock}.
However, it is necessary to keep in mind
that the extraction of the value of $ G_M^n $
from the data in this region of $Q^2$
is based on a rather model dependent procedure.

The electromagnetic form factors of the nucleon
enter into the relations (\ref{E033}) for the asymmetries
through the ratios
$ R_{p(n)}(Q^2) $.
We have considered
these ratios for $Q^2$ values in the interval
$ 0.5 \le Q^2 \le 10 \, \mbox{GeV}^2 $.
In our calculations
we have used the latest parameterizations
of the electromagnetic form factors of the nucleon
\cite{WT2,Bosted}
that are aimed to describe all existing
data.
In order to calculate the ratios
$ R_{p(n)}(Q^2) $
with error bands,
we have also done our own fit of the data
on the nucleon form factors in the $Q^2$ region under consideration.
For the magnetic form factor of the proton we have taken
the following two-poles expression:
\begin{equation}
\frac{G_M^p}{\mu_p} =
\frac{a_1}{1+a_2 Q^2} +
\frac{1-a_1}{1+a_3 Q^2}
\;,
\label{D01}
\end{equation}
which was proposed in Ref.\cite{Sofia}.
Taking into account possible violations of scaling,
we have considered the following parameterization
for the magnetic form factor of the neutron:
\begin{equation}
\frac{G_M^n}{\mu_n} =
\frac{G_M^p}{\mu_p} \left(1 + a_4 Q^2\right)
\;.
\label{D02}
\end{equation}
In Eqs.(\ref{D01}) and (\ref{D02})
$ a_1 , a_2 , a_3 , a_4 $
are variable parameters.
{}From the fit of the data
in the range
$ 0.5 \le Q^2 \le 10  \, \mbox{GeV}^2 $
(114 proton points
and 22 neutron points)
we have obtained the following
values for the parameters:
\begin{equation}
\begin{array}{l} \displaystyle
a_1 = -0.50 \pm 0.04 \;,
\\ \displaystyle
a_2 = 0.71 \pm 0.02 \;,
\\ \displaystyle
a_3 = 2.20 \pm 0.04 \;,
\\ \displaystyle
a_4 = -0.019 \pm 0.004 \;,
\end{array}
\label{D02P}
\end{equation}
with $\chi^2$/NDF=163/132,
which corresponds to a confidence level of 3.5\%.
The magnetic form factors of the proton and neutron
given by Eqs.(\ref{D01})--(\ref{D02P})
are presented in Figs.\ref{FIG1} and \ref{FIG2} (solid line).
In these figures the experimental data
(circles with error bar)
and
the form factors calculated with
the parameterizations of Refs.\cite{WT2} and \cite{Bosted}
are also presented (dashed and dot-dashed lines, respectively).
It is seen from Figs.\ref{FIG1} and \ref{FIG2}
that in the considered region of $Q^2$
there is a good agreement between
the different parameterizations of the
form factors and the experimental data.

It is obvious that the parameterization given
in Eqs.(\ref{D01})--(\ref{D02P}) is valid only in the limited
$Q^2$ region in which we have performed the fit.
We have also tried the parameterization
\begin{equation}
\frac{G_M^n}{\mu_n} =
\frac{G_M^p}{\mu_p}
\,
\frac{1 + a_4 Q^2}{1 + a_5 Q^2}
\;,
\label{D03}
\end{equation}
that provide
the same behaviour for
$G_M^p$ and $G_M^n$
at high $Q^2$.
However,
the accuracy of the neutron data
do not allow to determine the value of the parameter
$a_5$.

The results of our
calculations for the ratios
$ R_{p}(Q^2) $ and $ R_{n}(Q^2) $
are presented in
Figs.\ref{FIG3} and \ref{FIG4},
respectively.
The error bands correspond
to a confidence level of 90\%.
In the same figures
we have shown also the values of the ratios
$ R_{p}(Q^2) $ and $ R_{n}(Q^2) $
calculated with the parameterizations of the form factors given in
Refs.\cite{WT2} and \cite{Bosted}
(dashed and dot-dashed lines, respectively).
The corresponding curves are contained within the
above mentioned
error bands,
except in the region
$ Q^2 \lesssim 2 \, \mathrm{GeV}^2 $.
The error band that we have obtained
for the ratios
$ R_{p}(Q^2) $ and $ R_{n}(Q^2) $
is mainly due to the errors of the data on the neutron form factor.
It is clear that better neutron data are needed.

Now we will discuss the strange form factors of nucleon.
Some informations about the value
of $g_A^s$
were obtained
from the analysis of the BNL data
\cite{Ahrens}
on NC elastic scattering of neutrinos and antineutrinos
on protons in the range
$ 0.4 \le Q^2 \le 1.1 \, \mathrm{GeV}^2 $.
Under the assumptions
that the contribution of the vector strange form factors can be neglected
and that the axial strange form factor has the same dipole $Q^2$ behaviour
as the non-strange one,
the authors of Ref.\cite{Ahrens}
found
$ g_A^s = - 0.15 \pm 0.09 $,
which is compatible with the value given in Eq.(\ref{E006}).
The data from BNL were reanalyzed
in Ref.\cite{Garvey}.
In this paper the contributions of the strange axial
as well as
vector form factors were taken into account.
Dipole formulas were assumed
for both the vector and the axial strange form factors.
The fitting parameters were
$g_A^s$,
$F_M^s(0)=\mu_s$,
$ \displaystyle
\left\langle r_s^2 \right\rangle
=
-6\left[\frac{d F_V^s(Q^2)}{d Q^2}\right]_{Q^2=0}
$
and
$M_A$ (the axial cutoff mass).
The following
values for the parameters were obtained
in their fit III
(with $\chi^2$/NDF=9.28/11,
corresponding to 60\% CL):
$ g_A^s = -0.13 \pm 0.09 $,
$ \mu_s = -0.39 \pm 0.70 $,
$ \left\langle r_s^2 \right\rangle = - 0.11 \pm 0.16 \, \mathrm{fm}^2 $,
$ M_A = 1.049 \pm 0.023 \, \mathrm{GeV} $.
They found,
however,
strong correlations between
the values of $M_A$ and $g_A^s$
obtained from the fit.
If
$ M_A = 1.086 \pm 0.015 $,
(which is the value that was obtained in the BNL
experiment \cite{Ahrens}),
the data are well described
under the assumption that
$ g_A^s = \mu_s = \left\langle r_s^2 \right\rangle = 0 $
(with $\chi^2$/NDF=14.12/14,
corresponding to 44\% CL).

The strange vector form factors
at small $Q^2$ were considered by Jaffe in Ref.\cite{Jaffe}.
This calculation is based on the vector dominance model (VDM).
The contributions of $\omega$, $\phi$ and one heavier meson were
taken into account.
A fit
\cite{Hohler}
of the nucleon isoscalar form factors was used
and $\omega$--$\phi$ mixing was considered.
Large values for the strange matrix elements were obtained
in this work
($ -0.43 \le \mu_s \le -0.25 $,
$0.11 \le \left\langle r_s^2 \right\rangle \le 0.22$).
The axial strange form factor was also calculated
in the framework of VDM \cite{Kirch}
taking into account the
contributions of the isoscalar
axial vector mesons $f_1(1285)$ and $f_1(1420)$.

Let us also notice that in accordance with
the quark counting rule at high $Q^2$,
the axial and vector strange
form factors must decrease more rapidly than the non-strange
form factors:
\begin{equation}
F_i^s(Q^2) \sim \left(\frac{1}{Q^2}\right)^4
\label{D05}
\end{equation}
A $Q^2$ dependence of the strange form factors
stronger than the dipole one
was in fact considered in Refs.\cite{Musolf,Donn}
in analogy with the parameterization of
the electric form factor of the neutron
proposed in Ref.\cite{Galster}.

We have calculated the effect of the strange form factors
on the asymmetries $\cal{A}_{p(n)}$
determined by the relation (\ref{E011})
assuming some simple models.
We would like to stress that
these
calculations can be considered only as illustrations
of the possible effects of the strange form factors.
We expect that some real information about the values
of the strange form factors of the nucleon will be obtained
in future neutrino experiments.

In Figs.\ref{FIG5} and \ref{FIG6} the
results of the calculation of
quantities
$4|V_{ud}|^2{\cal A}_p$
and
$4|V_{ud}|^2{\cal A}_n$
as functions of $Q^2$
are shown.
Here we have assumed
that the axial and magnetic strange form factors are
given by the dipole formulas
\begin{equation}
F_A^s(Q^2) = \frac{g_A^s}{\left(1+
\displaystyle{\frac{Q^2}{{M_A^s}^2}}\right)^2}
\;,
\label{D08}
\end{equation}
\begin{equation}
G_M^s(Q^2) = \frac{\mu^s}{\left(1+
\displaystyle{\frac{Q^2}{{M_V^s}^2}}\right)^2}
\;,
\label{D09}
\end{equation}
where
$ M_A^s = M_A $
and
$ M_V^s = M_V $,
with the typical values
$ M_A = 1.032 \, \mathrm{GeV} $
and
$ M_V = 0.84 \, \mathrm{GeV} $.
The shadowed areas in Figs.\ref{FIG5} and \ref{FIG6}
are due to the
uncertainty of the electromagnetic form factors.
These areas have been
obtained under the assumption that $g_A^s=\mu_s=0$.
All the curves in Figs.\ref{FIG5} and \ref{FIG6}
have been obtained
with $g_A^s=-0.15$.
The dashed and dotted curves
display the effect of
axial strange form factor:
they were obtained
with $\mu_s=0$
utilizing our fit and the WT2 fit,
respectively,
for the magnetic form factors of the nucleon.
The solid and dot-dashed lines
show the effect of the contribution of
both the strange axial and vector form factors:
they were obtained
with $\mu_s=-0.3$
and
utilizing our fit and the WT2 fit,
respectively,
for the magnetic form factors of the nucleon.
The slow decrease of the asymmetry with $Q^2$ is due to the deviation of
the electromagnetic form factors from the dipole behaviour.
It is seen from Figs.\ref{FIG5} and \ref{FIG6}
that the contribution of the axial strange form factor
to the asymmetries
can be relatively
large if the constant
$g_A^s$ has the value given by the EMC and other data.
The combined effect of
the axial and vector strange form factors
depends on the signs of $g_A^s$ and $\mu_s$.
If the signs of these
constants are the same
(as we have assumed),
the contribution of the axial and vector strange form factors to the asymmetry
add up.
If the signs are opposite
(as it is the case
for example in Ref.\cite{Weigel}),
the contribution of
the vector strange form factor will compensate the contribution
of the axial one.

Until now
we considered
a dipole-behaviour of the strange form factors.
In the following we will calculate some
values of the asymmetries
$ \cal{A}_{p(n)} $
in the case of a $Q^2$-behaviour of the strange form factors
weaker and stronger than the dipole.
In Fig.\ref{FIG7} we present some calculations
of the quantity
$ R_p \cal{A}_p $
as a function of $Q^2$
in the case of a high-$Q^2$ decrease
of the strange form factors
stronger than the dipole.
For the strange form factors we have adopted
the following formulas:
\begin{eqnarray}
G_M^S(Q^2)
&=&
{\displaystyle
\mu_s
\over\displaystyle
\left( 1+ \frac{Q^2}{M_V^2}\right)^2
}
\,
{\displaystyle
1
\over\displaystyle
\left(1+\lambda_M^s \frac{Q^2}{4M^2}\right)
}
\;,
\label{D06}\\
F_A^s(Q^2)
&=&
{\displaystyle
g_A^s
\over\displaystyle
\left( 1+ \frac{Q^2}{M_A^2}\right)^2
}
\,
{\displaystyle
1
\over\displaystyle
\left(1+\lambda_A^s \frac{Q^2}{4M^2}\right)
}
\;,
\label{D07}
 \end{eqnarray}
where
$\lambda_A^s$ and $\lambda_M^s$
are unknown parameters.
The dotted and dot-dashed curves in Fig.\ref{FIG7}
have been calculated with
$\lambda_A^s=\lambda_M^s=3$.
The dotted and dashed line were calculated with
$g_A^s= -0.15$, $\mu_s=0$,
whereas
the dot-dashed and solid line with
$g_A^s=-0.15$, $\mu_s=-0.3$.
As it is seen from Fig.\ref{FIG7},
the contributions of the strange form factors
to the asymmetry tend to disappear at high $Q^2$
but could be relatively large in low $Q^2$ region
(at
$ Q^2 \lesssim 5 \, \mathrm{GeV}^2 $
in the cases considered).

Finally,
let us consider
the case of a high-$Q^2$ decrease
of the strange form factors
weaker than the dipole.
In order to improve the large $Q^2$ behaviour of the strange vector
form factors
proposed by Jaffe
\cite{Jaffe}
we added one additional pole to his expressions for
$F_V^s(Q^2)$ and $F_M^s(Q^2)$ :
\begin{equation}
F_V^s(Q^2)=Q^2\left[
\frac{a_\omega}{Q^2 + m_\omega^2} +
\frac{a_\phi}{Q^2 + m_\phi^2} +
\frac{a_1}{Q^2 + m_1^2} +
\frac{a_2}{Q^2 + m_2^2} \right]
\;,
\label{D010}
\end{equation}
\begin{equation}
F_M^s(Q^2)=
\frac{b_\omega m_\omega^2}{Q^2 + m_\omega^2} +
\frac{b_\phi m_\phi^2}{Q^2 + m_\phi^2} +
\frac{b_1 m_1^2}{Q^2 + m_1^2} +
\frac{b_2 m_2^2}{Q^2 + m_2^2}
\;.
\label{D011}
\end{equation}
We have taken the values of the parameters that characterize
the $\omega$ and $\phi$
poles and the third mass $m_1$
as in fit 8.1 of Ref.\cite{Jaffe}.
The value of the fourth mass was assumed to be
$ m_2 = 1.6 \, \mathrm{GeV} $
(which corresponds to a $\omega$ resonance).
The other parameters in
Eqs.(\ref{D010}) and (\ref{D011})
were fixed by imposing
the following asymptotic conditions:
\begin{equation}
\arraycolsep=0cm
\begin{array}{rcl} \displaystyle
F_V^s(Q^2) @>>{Q^2\to\infty}> 0
\null & \null \hspace{2cm} \null & \null \displaystyle
Q^2 F_M^s(Q^2) @>>{Q^2\to\infty}> 0
\\ \displaystyle
Q^2 F_V^s(Q^2) @>>{Q^2\to\infty}> 0
\null & \null \hspace{2cm} \null & \null \displaystyle
Q^4 F_M^s(Q^2) @>>{Q^2\to\infty}> 0
\end{array}
\label{D012}
\end{equation}
At $Q^2=0$
we have obtained
$ \left\langle r_s^2 \right\rangle = 0.055 \, \mathrm{fm}^2 $
and
$ \mu_s = -0.14 $,
to be compared with the values
$ \left\langle r_s^2 \right\rangle = 0.11 \, \mathrm{fm}^2 $
and
$ \mu_s = -0.25 $
of Ref.\cite{Jaffe}.

For the axial strange form factor we have employed a dipole
formula with a cutoff mass twice as large as $M_A$,
which corresponds
to a distribution of strange quarks inside of the nucleon
narrower than
the distributions of the $u$ and $d$ quarks
(see Refs.\cite{Beck,Kirch}).

The results of our calculations of
the quantity
$ R_p {\cal A}_p $
are shown in Fig.\ref{FIG8}.
The increase of the asymmetry with $Q^2$
is due to the fact that
the decrease of the strange form factors
at high $Q^2$ is slower than that
of the non-strange ones.
Taking into account the fact that
the asymptotical conditions for the strange form factors
are more stringent than those for
the non-strange form factors,
we would like to warn
that the behavior of the asymmetry
presented in Fig.\ref{FIG8} is strongly model dependent
(at least at high $Q^2$).

In conclusion,
we would like to make the following remark.
In most neutrino experiments
the interactions of neutrinos with
nuclei are investigated.
In order to obtain from these
experiments some informations about
the strange form factors of the nucleon,
the nuclear effects must be taken into account.
In the region of
relatively large $Q^2$
($Q^2 \gtrsim 0.5 \, \mathrm{GeV}^2 $)
in which we are interested,
the relativistic Fermi gas approximation
\cite{Horowitz}
can be used.
It was shown in Ref.\cite{Donn}
that in the case of quasi-elastic
electron-nuclei scattering this approximation works
reasonably well.
We plan to consider in detail the nuclear
effects in our future work.

\acknowledgments

It is a pleasure to thank
M. Anselmino,
J. Bernabeu,
A. Marchionni
and
A. Pullia
for fruitful discussions.


\begin{figure}[p]
\protect\caption{
Experimental data
\protect\cite{Data_p,Bartel}
of the proton magnetic form factor
scaled by the dipole fit
in the range
$ 0.5 < Q^2 < 10 \, \mbox{GeV}^2 $
(circles with error bar).
The fits given in
Ref.\protect\cite{WT2} (dashed line),
Ref.\protect\cite{Bosted} (dot-dashed line)
are shown
together with our fit (solid line).
}
\label{FIG1}
\end{figure}

\begin{figure}[p]
\protect\caption{
Experimental data
\protect\cite{Bartel,Hanson,Lung,Rock}
of the neutron magnetic form factor
scaled by the dipole fit
in the range
$ 0.5 < Q^2 < 10 \, \mbox{GeV}^2 $
(circles with error bar).
The fits given in
Ref.\protect\cite{WT2} (dashed line),
Ref.\protect\cite{Bosted} (dot-dashed line)
are shown
together with our fit (solid line).
}
\label{FIG2}
\end{figure}

\begin{figure}[p]
\protect\caption{
The value of the ratio $R_p$
obtained with our fit
of the proton and neutron magnetic form factors
(solid line).
The 90\% confidence level region of $R_p$
is depicted a shadowed band.
The dashed and dot-dashed lines
represent
the values of $R_p$
obtained with the fits given in
Ref.\protect\cite{WT2} (WT2),
Ref.\protect\cite{Bosted} (Bosted),
respectively.
The dotted line
represent the constant value of $R_p$ in the scaling
approximation.
}
\label{FIG3}
\end{figure}

\begin{figure}[p]
\protect\caption{
The value of the ratio $R_n$
obtained with our fit
of the proton and neutron magnetic form factors
(solid line).
The 90\% confidence level region of $R_n$
is depicted a shadowed band.
The dashed and dot-dashed lines
represent
the values of $R_n$
obtained with the fits given in
Ref.\protect\cite{WT2} (WT2),
Ref.\protect\cite{Bosted} (Bosted),
respectively.
The dotted line
represent the constant value of $R_n$ in the scaling
approximation.
}
\label{FIG4}
\end{figure}

\begin{figure}[p]
\protect\caption{
Plot of
$4|V_{ud}|^2{\cal A}_p$
as a function of $Q^2$.
The shadowed area corresponds to the
uncertainty induced by the error of $R_p$ (see Fig.\ref{FIG3})
in the absence of strange contributions.
All the curves were obtained using
a dipole form for
$ F_A^s(Q^2) $
with
$g_A^s=-0.15$
and
$M_A^s=M_A$.
The dashed (dotted) curve was obtained
with $ G_M^S(Q^2) = 0 $
utilizing our fit (the WT2 fit)
for the magnetic form factors of the nucleon.
The solid (dot-dashed) line
was obtained using
a dipole form for
$ G_M^S(Q^2) $
with $\mu_s=-0.3$, $M_V^s=M_V$
and
utilizing our fit (the WT2 fit)
for the magnetic form factors of the nucleon.
}
\label{FIG5}
\end{figure}

\begin{figure}[p]
\protect\caption{
Plot of
$4|V_{ud}|^2{\cal A}_n$
as a function of $Q^2$.
The shadowed area corresponds to the
uncertainty induced by the error of $R_n$ (see Fig.\ref{FIG4})
in the absence of strange contributions.
All the curves were obtained using
a dipole form for
$ F_A^s(Q^2) $
with
$g_A^s=-0.15$
and
$M_A^s=M_A$.
The dashed (dotted) curve was obtained
with $ G_M^S(Q^2) = 0 $
utilizing our fit (the WT2 fit)
for the magnetic form factors of the nucleon.
The solid (dot-dashed) line
was obtained using
a dipole form for
$ G_M^S(Q^2) $
with $\mu_s=-0.3$, $M_V^s=M_V$
and
utilizing our fit (the WT2 fit)
for the magnetic form factors of the nucleon.
}
\label{FIG6}
\end{figure}

\begin{figure}[p]
\protect\caption{
Plot of
$ R_p {\cal A}_p $
as a function of $Q^2$.
As in Fig.\ref{FIG5},
the shadowed area corresponds to the
uncertainty induced by the error of $R_p$ (see Fig.\ref{FIG3})
in the absence of strange contributions.
The dashed and solid lines
coincide with the corresponding lines
of Fig.\ref{FIG5}.
The dotted and dot-dashed curves
were obtained with
the Galster parameterization
(\ref{D06}) and (\ref{D07})
for
$ G_M^s(Q^2) $ and $ F_A^s(Q^2) $.
The values of the parameters are:
$g_A^s=-0.15$,
$M_A^s=M_A$,
$\lambda_A^s=\lambda_M^s=3$,
$M_V^s=M_V$,
$\mu_s=0$ and $\mu_s=-0.3$
for the dotted and dot-dashed curve,
respectively.
}
\label{FIG7}
\end{figure}

\begin{figure}[p]
\protect\caption{
Plot of
$ R_p {\cal A}_p $
as a function of $Q^2$.
As in Figs.\ref{FIG5} and \ref{FIG7},
the shadowed area corresponds to the
uncertainty induced by the error of $R_p$ (see Fig.\ref{FIG3})
in the absence of strange contributions.
All the curves were obtained using
a dipole form for
$ F_A^s(Q^2) $
with
$g_A^s= -0.15$.
The dashed line,
which was obtained with
$ G_M^S(Q^2) = 0 $
and
$ M_A^s = M_A $,
coincide with the corresponding line
of Figs.\ref{FIG5} and \ref{FIG7}.
The dotted line was obtained with
$ G_M^S(Q^2) = 0 $
and
$ M_A^s = 2 M_A $.
The solid (dot-dashed) curve
was obtained with
the modified Jaffe parameterization for $ G_M^S(Q^2) $
and
$ M_A^s = M_A $ ($ M_A^s = 2 M_A $).
}
\label{FIG8}
\end{figure}

\newpage
\setcounter{figure}{0}

\begin{figure}[p]
\begin{center}
\mbox{\epsfig{file=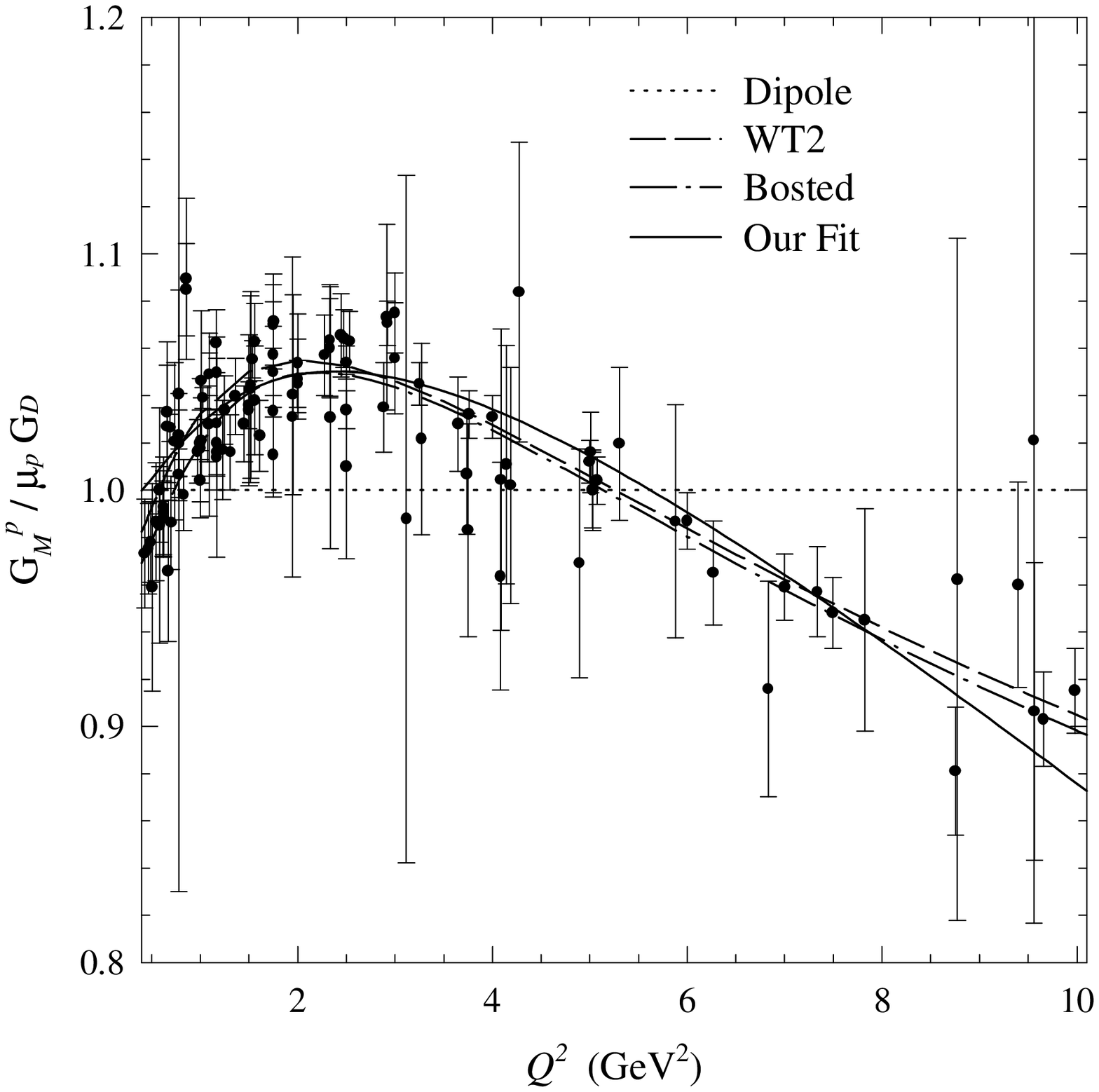,width=\textwidth}}
\end{center}
\vspace{1cm}
\begin{center}
{\Large Figure \ref{FIG1}}
\end{center}
\end{figure}

\begin{figure}[p]
\begin{center}
\mbox{\epsfig{file=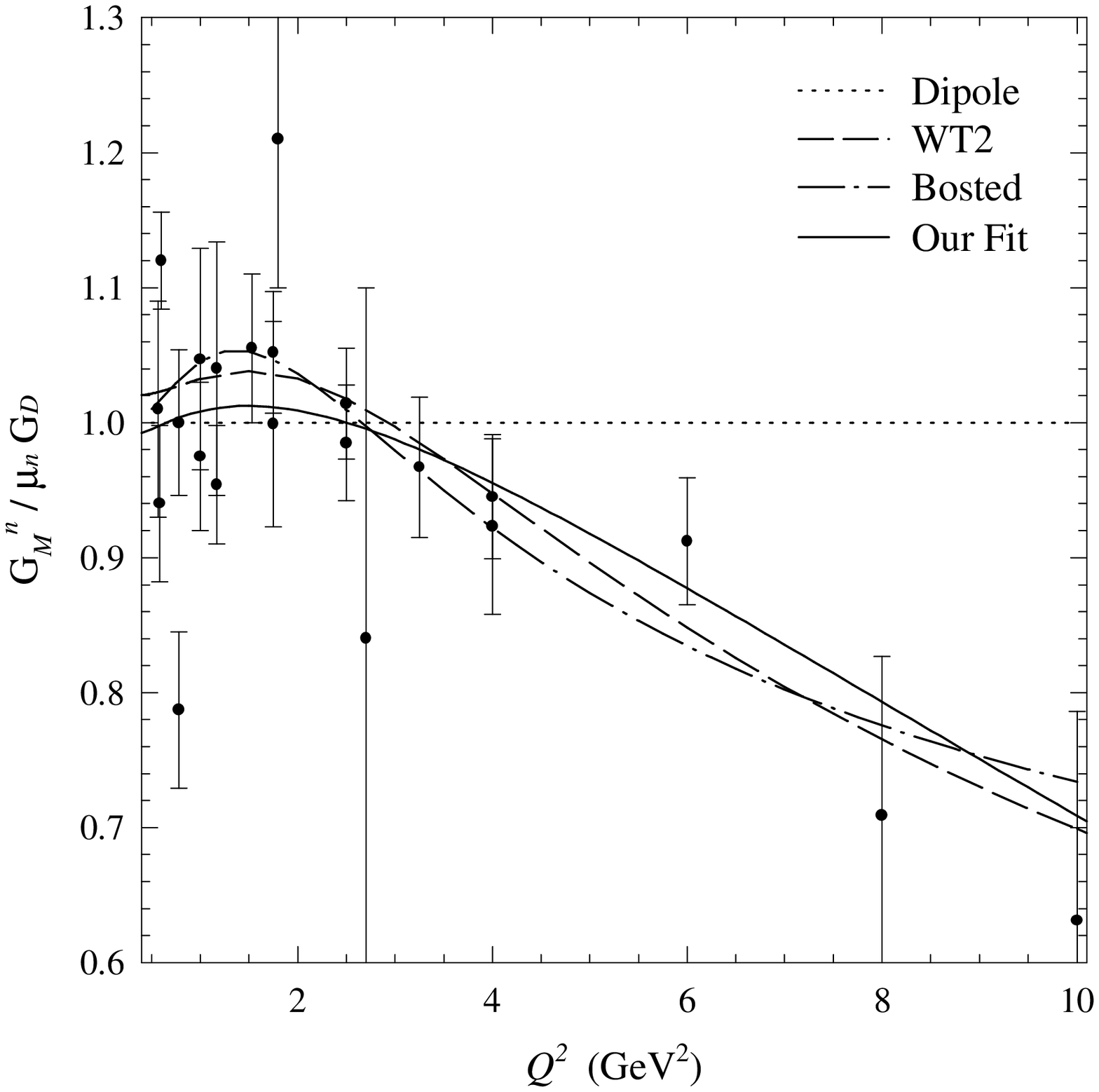,width=\textwidth}}
\end{center}
\vspace{1cm}
\begin{center}
{\Large Figure \ref{FIG2}}
\end{center}
\end{figure}

\begin{figure}[p]
\begin{center}
\mbox{\epsfig{file=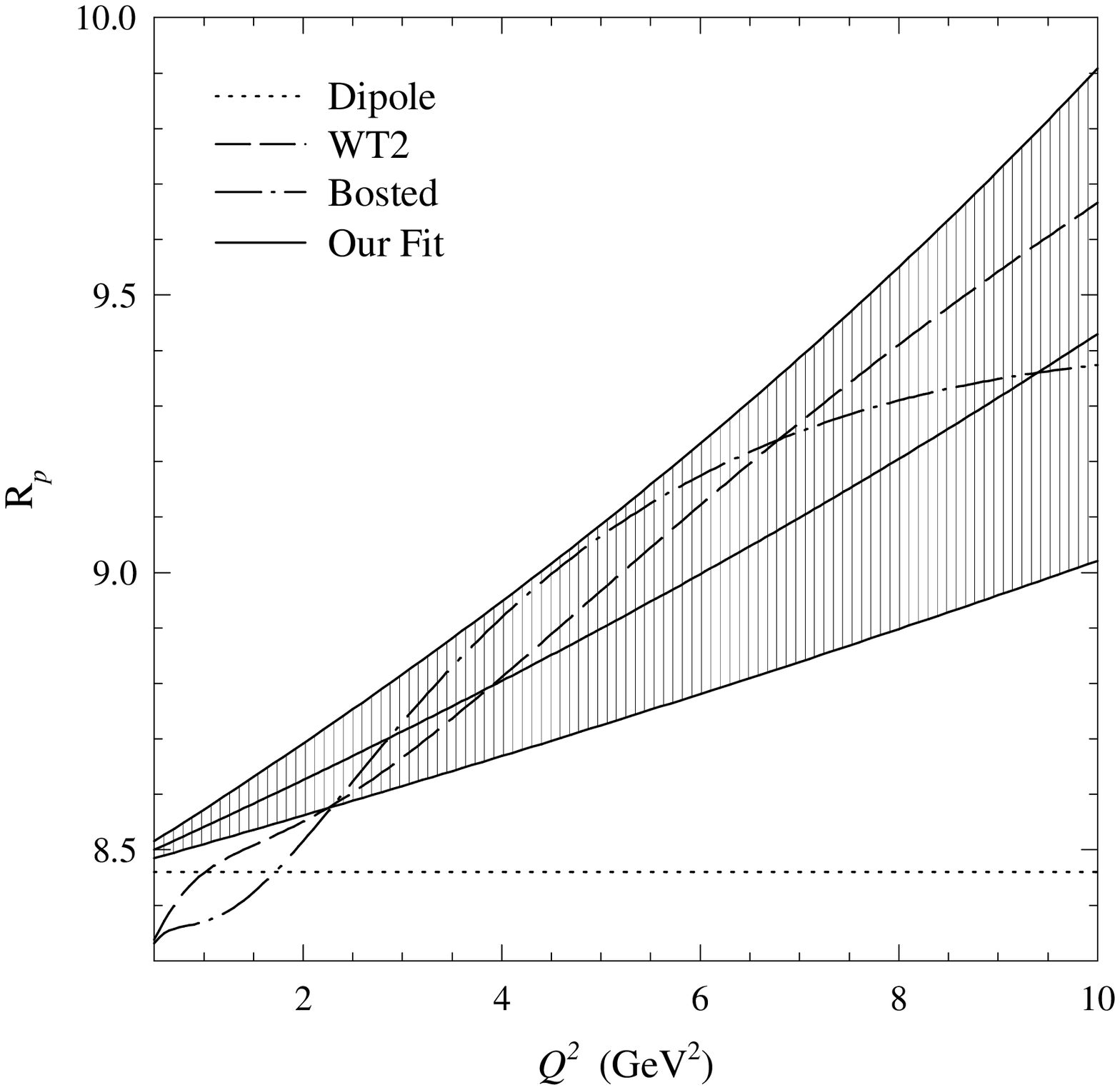,width=\textwidth}}
\end{center}
\vspace{1cm}
\begin{center}
{\Large Figure \ref{FIG3}}
\end{center}
\end{figure}

\begin{figure}[p]
\begin{center}
\mbox{\epsfig{file=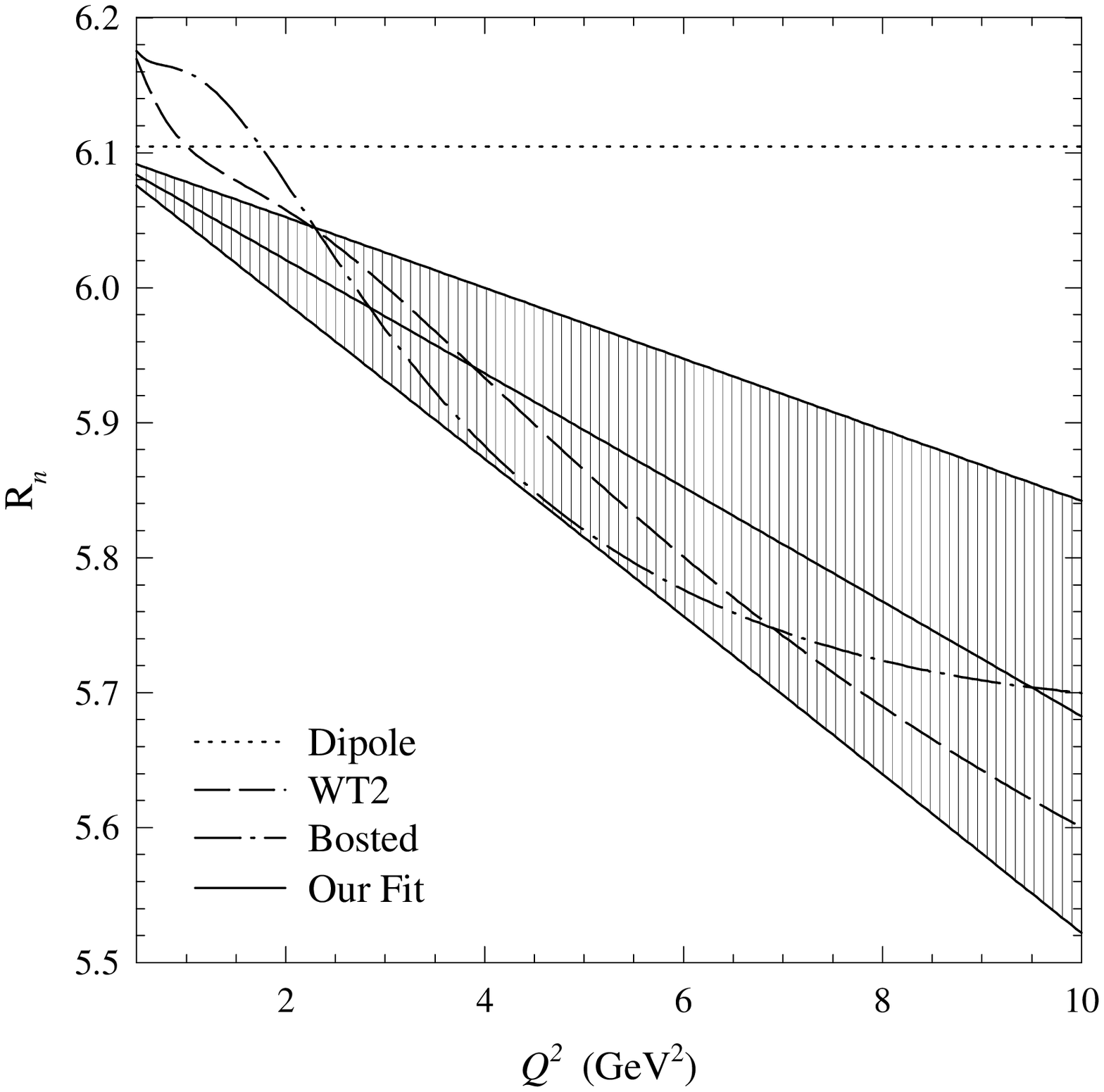,width=\textwidth}}
\end{center}
\vspace{1cm}
\begin{center}
{\Large Figure \ref{FIG4}}
\end{center}
\end{figure}

\begin{figure}[p]
\begin{center}
\mbox{\epsfig{file=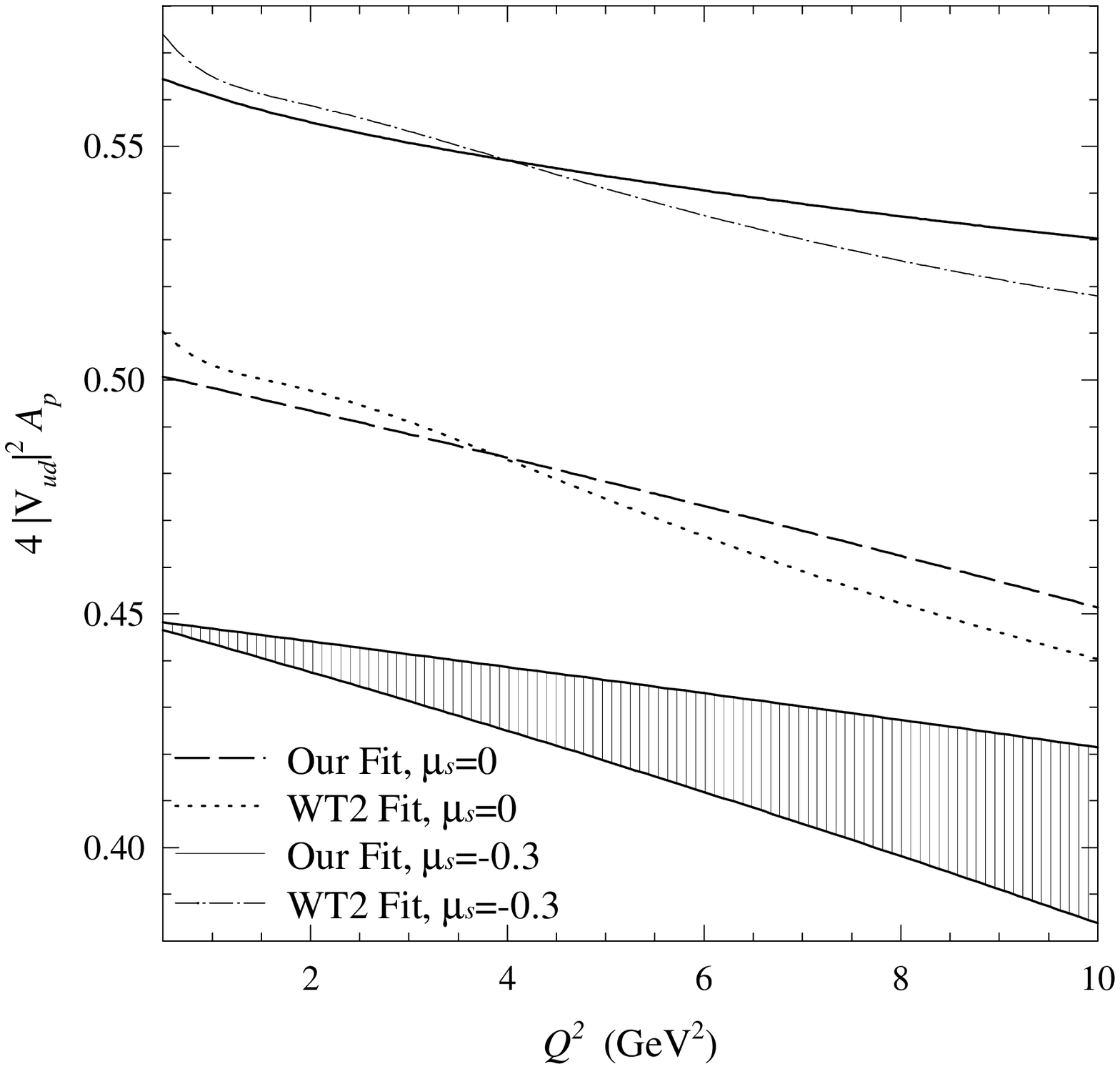,width=\textwidth}}
\end{center}
\vspace{1cm}
\begin{center}
{\Large Figure \ref{FIG5}}
\end{center}
\end{figure}

\begin{figure}[p]
\begin{center}
\mbox{\epsfig{file=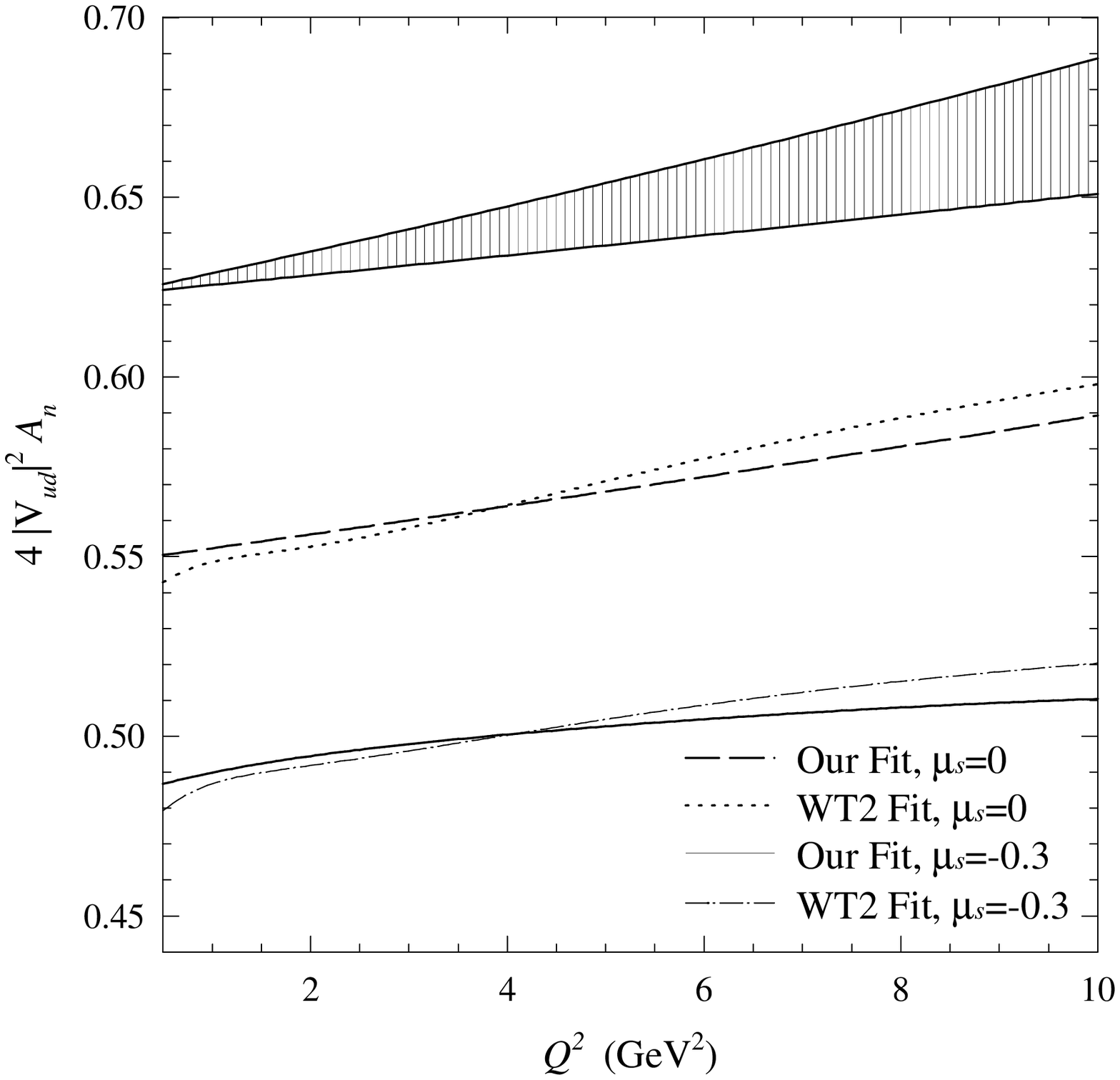,width=\textwidth}}
\end{center}
\vspace{1cm}
\begin{center}
{\Large Figure \ref{FIG6}}
\end{center}
\end{figure}

\begin{figure}[p]
\begin{center}
\mbox{\epsfig{file=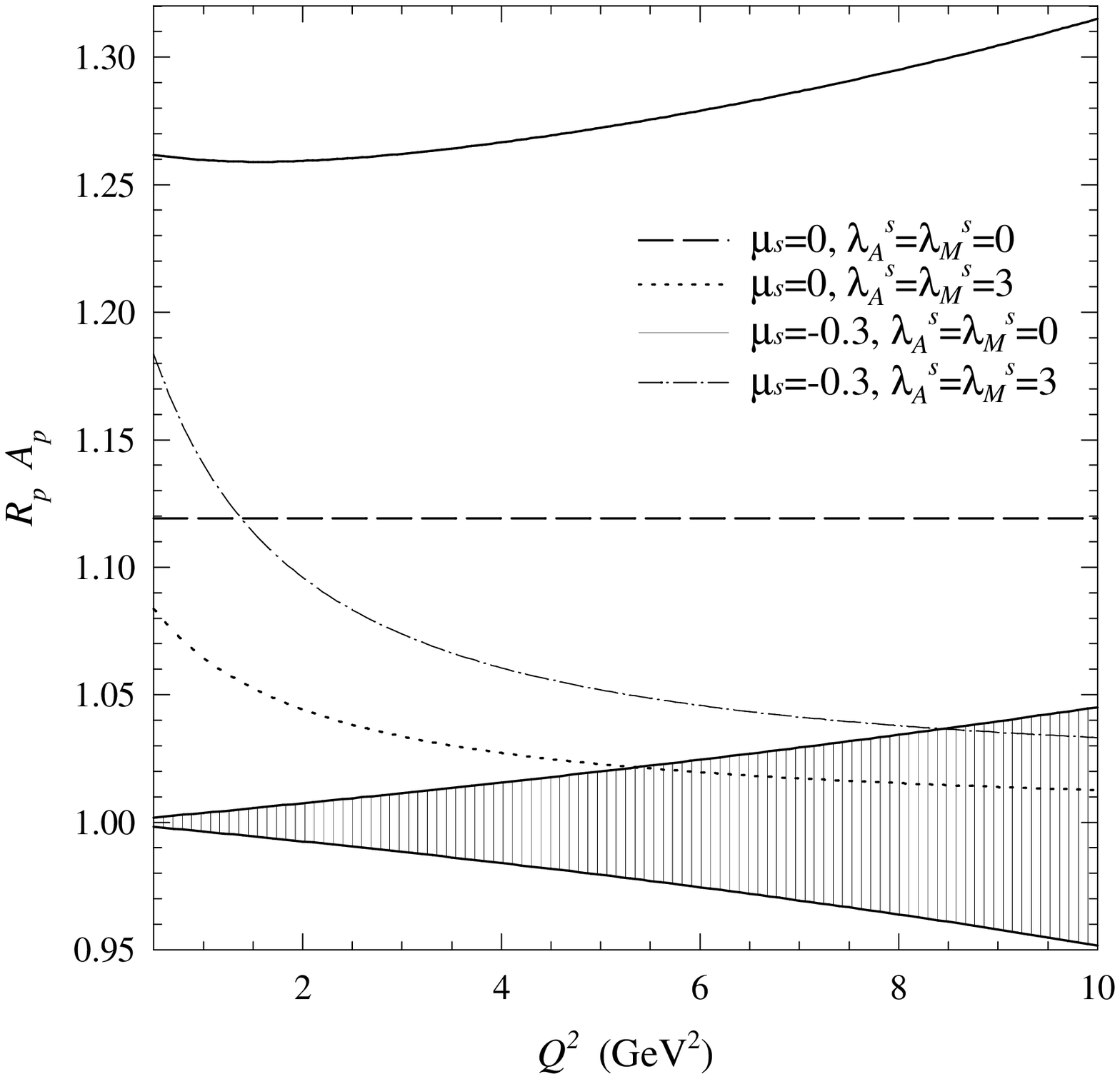,width=\textwidth}}
\end{center}
\vspace{1cm}
\begin{center}
{\Large Figure \ref{FIG7}}
\end{center}
\end{figure}

\begin{figure}[p]
\begin{center}
\mbox{\epsfig{file=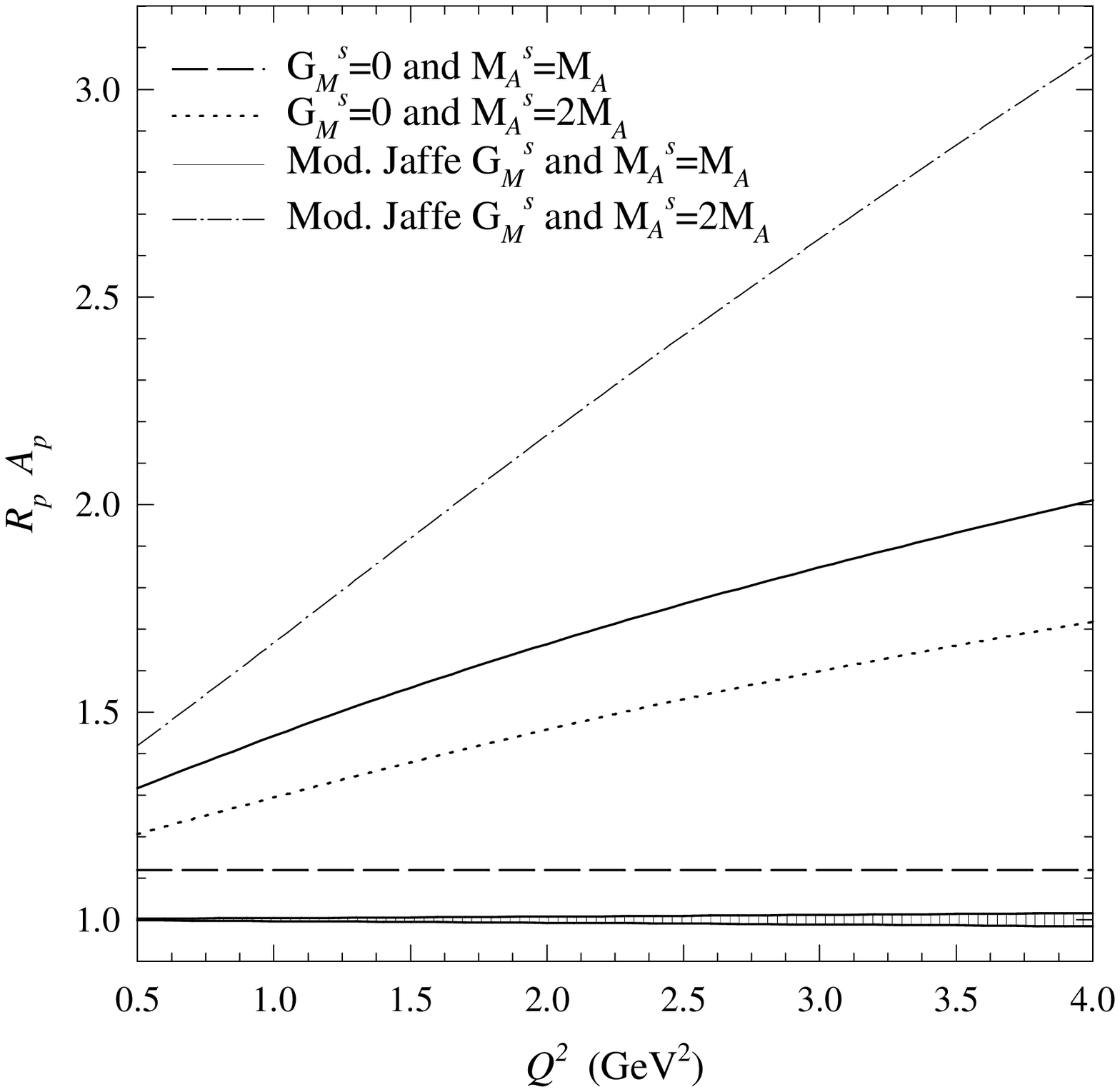,width=\textwidth}}
\end{center}
\vspace{1cm}
\begin{center}
{\Large Figure \ref{FIG8}}
\end{center}
\end{figure}


\begin{references}

\bibitem{EMC}
J. Ashman {\it et al.},
Phys. Lett. B {\bf 206}, 364 (1988);
Nucl. Phys B {\bf 328},1 (1989).

\bibitem{ANSELMINO}
M. Anselmino, A. Efremov and E. Leader,
preprint CERN-TH/7216/94
(hep-ph\-@@\-xxx.\-lanl.\-gov/\-9501369), April 1994,
to appear in Phys. Rep.

\bibitem{CERN}
D. Adams et al.,
Phys. Lett. B {\bf 329}, 399 (1994).

\bibitem{SLAC}
K. Abe et al.,
Phys. Rev. Lett. {\bf 74}, 346 (1995).

\bibitem{QCDSUMRULE}
J. Kodaira {\it et al.},
Nucl. Phys. B {\bf 159}, 99 (1979);
Nucl. Phys. B {\bf 165}, 129 (1979);
Phys. Rev. D {\bf 20}, 627 (1979).

\bibitem{CLOSE}
F. Close and R. Roberts,
Phys. Lett. B {\bf 316}, 165 (1993).

\bibitem{Ellis}
J. Ellis and M. Karliner, Phys. Lett. B {\bf 341}, 397 (1995).

\bibitem{ANOMALY}
A.V. Efremov and O.V. Teryaev,
JINR Report No. E2-88-287 (1988), unpublished;
G. Altarelli and G.G. Ross, Phys. Lett. B {\bf 212},391 (1988).

\bibitem{Landshoff}
S.D. Bass an P.V. Landshoff, Phys. Lett. B {\bf 336}, 537 (1994).

\bibitem{SU3}
J. Lichtenstadt and H.J. Lipkin,
preprint TAUP-2244-95
(hep-ph\-@@\-xxx.\-lanl.\-gov/\-9504277).

\bibitem{NEUTRINOP}
D.B. Kaplan and A. Manohar, Nucl. Phys. B {\bf 310}, 527 (1988);
J. Ellis and M. Karliner, Phys. Lett. B {\bf 213}, 73 (1988).

\bibitem{Garvey}
G.T. Garvey, W.C. Louis and D.H. White,
Phys. Rev. C {\bf 48}, 761 (1993).

\bibitem{Ahrens}
L.A. Ahrens et al.,
Phys. Rev. D {\bf 35}, 785 (1987).

\bibitem{LBNO}
K. Nishikawa,
INS-Rep-924, April 1992;
A.K. Mann et al.,
BNL-PROPOSAL-889, Jan 1993;
MINOS Coll.,
NuMI note NUMI-L-63, February 1995;
NUMI-L-79, April 1995;
ICARUS Coll.,
Gran Sasso Lab. preprint LNGS-94/99-I,
May 1994.

\bibitem{RPP}
Review of Particle Properties,
Phys. Rev. D {\bf 50}, 1173 (1994).

\bibitem{Data_p}
L. Andivahis {\it et al.}, Phys. Rev. D {\bf 50}, 5491 (1994);
R.G. Arnold {\it et al.}, Phys. Rev. Lett. {\bf 57}, 174 (1986);
P.E. Bosted {\it et al}, Phys. Rev. C {\bf 42}, 38 (1990);
P.N. Kirk {\it et al.}, Phys. Rev. D {\bf 8}, 63 (1973);
D.Krupa {\it et al.}, J. Phys. G {\bf 10}, 455 (1984).

\bibitem{Bartel}
W. Bartel {\it et al.}, Nucl. Phys. B {\bf 58}, 429 (1973).

\bibitem{Hanson}
K.M. Hanson {\it et al.}, Phys. Rev. D {\bf 8}, 753 (1973).

\bibitem{Lung}
A. Lung {\it et al.}, Phys. Rev. Lett. {\bf 70}, 718 (1993).

\bibitem{Rock}
S. Rock {\it et al.}, Phys. Rev. Lett {\bf 49}, 1139 (1982).

\bibitem{WT2}
K. Watanabe and H. Takahashi,
Phys. Rev. D {\bf 51}, 1423 (1995).

\bibitem{Bosted}
P.E. Bosted,
Phys. Rev. C {\bf 51}, 509 (1995).

\bibitem{Sofia}
S.I. Bilen'kaya and Yu.M. Kazarinov,
Sov. J. Nucl. Phys. {\bf 32}, 382 (1980).

\bibitem{Jaffe}
R.L. Jaffe,
Phys. Lett. B {\bf 229}, 275 (1989).

\bibitem{Hohler}
G. H\"ohler et al.,
Nucl. Phys. B {\bf 224}, 505 (1976).

\bibitem{Beck}
D.H. Beck,
Phys. Rev. D {\bf 39}, 3248 (1989).

\bibitem{Kirch}
M. Kirchbach and H. Arenh\"ovel,
Contribution to the
{\it Int. Conf. on Physics with GeV-Particle Beams},
22-25 Aug 1994, J\"ulich, Germany
(hep-ph\-@@\-xxx.\-lanl.\-gov/\-9409293).

\bibitem{Musolf}
M.J. Musolf et al.,
Phys. Reports {\bf 239}, 1 (1994).

\bibitem{Donn}
T.W. Donnelly et al.,
Nucl. Phys. A {\bf 541}, 525 (1992).

\bibitem{Galster}
S. Galster,
Nucl. Phys. B {\bf 32}, 221 (1971).

\bibitem{Weigel}
H. Weigel et al.,
Phys. Lett. B {\bf 353}, 20 (1995).

\bibitem{Horowitz}
C.J. Horowitz et al.,
Phys. Rev. C {\bf 48}, 3078 (1993).

\end{references}
\end{document}